\documentclass[aps,prl,twocolumn,superscriptaddress]{revtex4-1}
\usepackage{xcolor}

\usepackage{graphicx}

\begin{document}


\title{Atomistic origin of metal versus charge-density-wave phase separation in indium atomic wires on Si(111)}



\author{Sun Kyu Song}
\affiliation{Center for Artificial Low Dimensional Electronic Systems, Institute for Basic Science (IBS), Pohang 37673, Republic of Korea} 

\author{Han Woong Yeom}
\email[]{yeom@postech.ac.kr}
\affiliation{Center for Artificial Low Dimensional Electronic Systems, Institute for Basic Science (IBS), Pohang 37673, Republic of Korea} 
\affiliation{Department of Physics, Pohang University of Science and Technology (POSTECH), Pohang 37673, Republic of Korea}



\begin{abstract}

We investigate in atomic scale the electronic phase separation occurring in the well known quasi 1D charge-density wave (CDW) phase of the In atomic wire array on a Si(111) surface.
The characteristic atomic scale defects, originated from excess In atoms, are found to be actively involved in the formation of the phase boundary between the metallic and the CDW phases by extensive analysis of scanning tunneling microscopy images at various temperatures. 
These particular defects flip the phase of the quasi 1D CDW to impose strong local constraints in the CDW correlation.  
We show that such local constraints and the substantial interwire CDW interaction induce local condensates of CDW and the phase separation between the metallic and the CDW phases. 
This work unveils the atomistic origin of the electronic phase separation, highlighting the importance of atomic scale structures of defects and their collective interaction in electronically inhomogeneous materials.

\end{abstract}



\maketitle



Electronic phase separation (EPS) is the static coexistence of distinct electronic phases  in a microscopic length scale at low temperature~\cite{SWCheong2002a}.
EPS has been recognized as an underlying mechanism of exotic electronic phenomena such as colossal magnetoresistance~\cite{MUehara1999a} and high-temperature superconductivity~\cite{JTPark2009a, EVLdeMello2009a}.
The investigation of EPS is not only crucial for understanding such exotic phenomena, but also necessary for manipulating competing electronic phases toward novel device applications.
Naturally, in order to understand the emergence of EPS, the lateral microscopic investigation of electronically distinct phases is crucial.
Indeed, microscopic studies have shed light on the origins of EPS in complex oxides such as strain~\cite{KHAhn2004,Fontcuberta2006a,Mukhopadhyay2008a}, charge doping~\cite{Sboychakov2009} or chemical disorder~\cite{Zhu2016a,TMiao202a}.
However, in spite of the important progress made by recent microscopic measurements~\cite{MUehara1999a,Lang2002,JTPark2009a,JTao2009a,Dhital2014,Du2015a,Zhu2016a}, the understanding of the atomic scale role of individual dopants and defects has been largely limited~\cite{GKinoda2003a}.
This is due partly to the lack of measurements using a proper spectroscopic probes with atomic scale resolution and partly to multiple degrees of freedom intertwined in EPS for complex materials.
For these reasons, the investigation of EPS in a simple model system with an atomic scale resolution can provide a new insight into EPS-related phenomena. 

In this context, we focus on the coexistence between a metallic state and an insulating charge-density-wave (CDW) state ~\cite{BSipos2008,HMorikawa2010a,TRitschel2013,YLiu2013,Zhang2014c,DCho2016a}.
The EPS in quasi 2D chalcogenide systems were reported to be induced by pressure~\cite{TRitschel2013}, disorder~\cite{PXu2010}, doping~\cite{YLiu2013} or external excitations~\cite{LStojchevska2014a,DCho2016a}, which has been extensively discussed for novel electronic phenomena~\cite{BSipos2008,DCho2016a} and functionalities~\cite{LStojchevska2014a,IVaskivskyi2016NC,MYoshida2015a}.
These systems, however, also have various degrees of freedom involved including charge, lattice, spin, and strong electron correlation. 
On the other hand, the formation of EPS in a unique quasi 1D CDW system of the In atomic wire array on Si(111) can be a simpler model system without any magnetic ordering and substantial electron correlation involved~\cite{Yeom1999k}.
The previous studies revealed that the existence of EPS in this system as stabilized by adsorbates~\cite{HMorikawa2010a} or vacancy defects~\cite{Zhang2014c}. 
However, the microscopic origin of EPS is not sufficiently clear since the previous studies did not provide detailed atomic scale connection between the phase landscape and individual adsorbates or vacancies.

In this Letter, we investigate the atomic scale phase boundaries formed on the electronically phase separated In/Si(111) system using variable-temperature scanning tunneling microscopy (STM).
We identify that two characteristic atomic scale defects, which are the most popular defects in this system, are actively involved in the phase boundaries.
Atomically-resolved STM images reveal that the local constraints on the CDW imposed by the characteristic atomic structure of the defects under the interwire CDW coupling hinders the formation of CDW domains, through which the electronic phase separation emerges.
The atomic scale mechanism of EPS in a quasi 1D CDW system is, thus, unambiguously pined down, which can be considered in various EPS systems based on charge orderings.
These results point out the importance of detailed atomic scale measurements of phase boundaries and defects in understanding inhomogeneous electronic systems.


Our experiments were performed by a commercial ultrahigh vacuum STM equipped with a Joule-Thomson type cryostat (SPECS, Germany) at between 78 and 120 K. 
An $n$-type flat Si(111) substrate was cleaned by flash heating to $1500~\mathrm{K}$ a few times.
To grow the defective In atomic wire array on Si(111), In atoms over the optimum coverage of one monolayer was thermally deposited on the clean Si(111)$7\times7$ surface kept at $600~\mathrm{K}$~\cite{Yeom1999k}.
After deposition, the Si(111)$4\times1$-In surface was confirmed with low energy electron diffraction and STM.
All STM images were obtained in the constant-current mode with the tunneling current of $I_t=100$~pA and the sample bias of $V_s=\pm0.5$~V for empty and filled states, respectively.


\begin{figure}[tb]
\includegraphics[width=8cm]{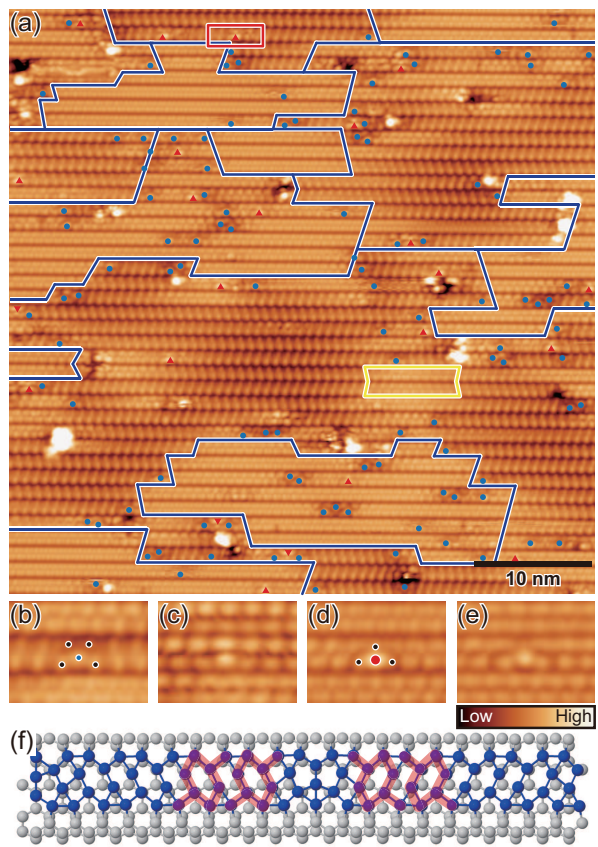}
\caption{
(a) STM topographic images ($V_s=-0.5$~V) of the electronic phase separation in the In/Si(111) surface taken at 78~K with its pristine CDW transition at 130 K. Blue lines indicate domain boundaries between the metallic (bright) and the insulating CDW (dark) phases.
The yellow box indicates two chiral solitons clustered. 
Blue dots and red triangles indicate characteristic the atomic-scale defects zoomed up in (b) and (d), respectively [their empty state images with $V_s=0.5$~V are shown in (c) and (e), respectively].
Dots indicate local structures of both defects for filled states.
(f) Atomic structure model of the 'M'-shaped defect shown in (b) and (c), which contains one excess In atom~\cite{SKSong2019a}. 
Blue (gray) balls indicates the indium (silicon) atoms.
Red lines indicates the In hexagon structure characteristic to the CDW state. 
}
\label{fig1}
\end{figure}


Figure~\ref{fig1}(a) shows the electronically phase-separated In atomic wire array on a Si(111) surface obtained at 78~K.
The metallic $4\times1$ (brighter regions) and the insulating $8\times2$ phase coexist well below the transition temperature of 130 K.
The difference of metallic and insulating domains is apparent even in the normal topographic measurement in filled states due to the clear structural difference between the two phases and the different topographic contrast reflecting the existence of the band gap for the insulating phase.
The atomically-resolved STM image reveals clearly a substantial density of defects in various forms such as bright adatom clusters, dark vacancies, and atomic scale defects (blue dots and red triangles).

The EPS occurs in samples prepared with an unoptimized In coverage on purpose.
In the present case, the key parameter is the density of the atomic scale defects, which is controlled by the In deposition or an high temperature post-annealing.
The atomic scale defects have better defined structures compared to clusters of vacancies or adatoms.
Two most well characterized and dominant atomic scale defects ('M' and 'A' shapes) are zoomed in Figs.~\ref{fig1}(b)$-$\ref{fig1}(e), respectively.
The population of these defects are $\sim69\%$ ('M') and $\sim17\%$ ('A') of all defects (Table.~S1), as counted in STM images covering a large area of 250~nm $\times$ 250~nm.
Note that about $90\%$ of the defects are observed near the domain boundaries or within the metallic $4\times1$ domains (Fig.~S1).
The other types of defects, the clusters of adatoms or vacancies are located rather randomly, as indicated by a much smaller number, $\sim41\%$, in a similar count.
It straightforwardly suggests that the EPS or the stabilization of the metallic phase below the CDW transition temperature is mainly related to the atomic scale defects.
In contrast, a previous STM study found that the EPS is enhanced by the extra annealing of the optimally prepared sample~\cite{Zhang2014c}. 
While this work focused on the large vacancies generated by the annealing, the direct relationship between local vacancy distributions and the domain landscape was not established at all.
Instead, the DFT calculations in the same work showed that the global strain can stabilize the metallic phase ~\cite{Zhang2014c}. 
However, it is not clear at all what kind of strain is actually imposed by the vacancies.
We would like to point out that such an annealing would produce not only vacancies but also In adatoms, ejected during the formation of vacancies. 
The role of atomic scale defects for the EPS in the present system will be made unambiguous by more detailed microscopic observations discussed below.

In fact, the major atomic scale defects ('M') in Figs.~\ref{fig1}(b) and \ref{fig1}(c) were already reported as the so-called short phase-flip defect (PFD) in the previous studies~\cite{Kim2012n,Lee2019a}.
These defects commonly force to flip the CDW unitcells along the wire with respect to an axis perpendicular to the wire due to their mirror symmetric structure.
The atomic origin of the short PFD is revealed as excess In adatoms experimentally~\cite{Lee2019a} and its atomic structure model was suggested by the DFT calculations~\cite{SKSong2019a, Lee2019a,Razzaq2020a} [see Fig.~\ref{fig1}(f)].
It is notable that both sides of this structure has almost the same atomic arrangement with one side of the indium hexagon of the CDW unitcell (see the guide lines in the figure). 
This arrangement and the mirror symmetric structure of the defect make possible the compact (just one atomic unitcell) matching with the neighboring CDW unitcells and the opposite CDW orientations on different sides of a defect, respectively. 
In terms of phonons, this defect pins the sheer phonon mode between the upper and the lower In chains within a wire, which is crucial in the CDW distortion~\cite{SWippermann2010a,SKSong2019a}.
The sheer phonon direction flips across this defect, which is the underlying mechanisms of the CDW phase pinning. 

In addition to this well known defect, the other atomic scale defect shown in Figs.~\ref{fig1}(d) and \ref{fig1}(e) has not been actively discussed before due probably to its small population.
In clear contrast to the above defect, the CDW pinning effect of this defect is limited. 
Some of these defects are observed with a local CDW phase flip but the others not (see Fig.~S2 in the supplementary materials). 
The origin of such a weak CDW phase pinning is not clear at present.
In the following discussion, we mainly focus on the strong PFDs.





\begin{figure}[tb]
\includegraphics[width=8cm]{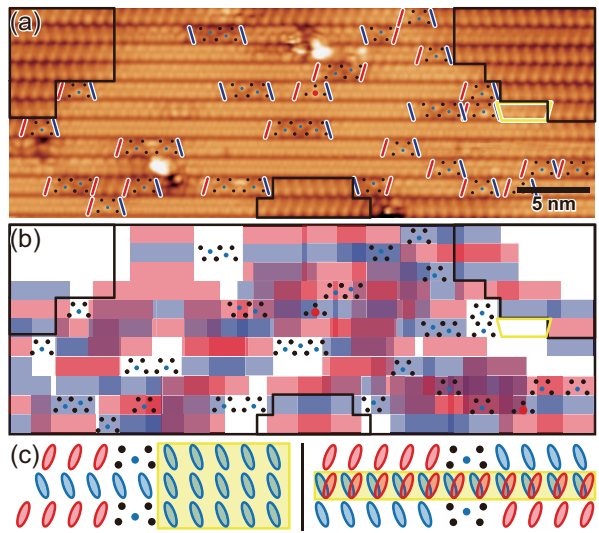}%
\caption{
(a) STM topographic images ($V_s=-0.5$~V) of a single $4\times1$ metallic domain and its domain boundaries with the $8\times2$ CDW phase (outlined by black lines) at 78~K.
Dots indicate local protrusions of the atomic scale defects (PFDs) as defined in Figs. 1(b) and 1(d). 
Red and blue bars denote the opposite CDW orientations imposed by the PFDs.
(b) Schematics of the frustrated CDW orientations due to the PFDs (indicated by the dots).
Red and blue stripes indicate the presumed phase coherent segments of within a single wire with its CDW orientations fixed by the PFDs, following the orientation of red and blue bars in (a).
(c) Schematics of frustrated CDW orderings to induce the electronic phase separation.
Red and blue ovals indicate the CDW orientations and yellow region indicates the interwire or intrawire disorder.
}
\label{fig2}
\end{figure}

\begin{figure}[tb]
\includegraphics[width=8cm]{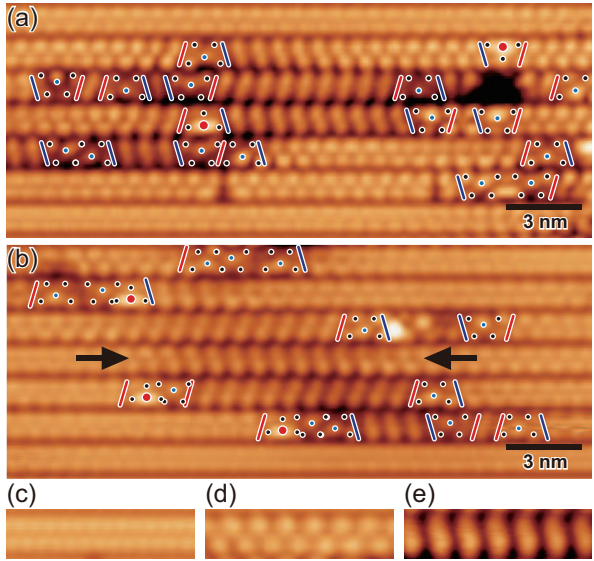}%
\caption{
(a), (b) STM topographic images ($V_s=-0.5$~V) of two small domains in the $8\times2$ CDW state surrounded by typical PFDs (dots) and metallic domains at 95~K.
Red and blue bars denote the CDW orientation imposed by defects.
(c), (d), (e) Close-up STM images of two metallic structures of $4\times1$ and $4\times2$ and the $8\times2$ CDW structures at 95~K, respectively.
}
\label{fig3}
\end{figure}

In considering the microscopic mechanism of the EPS induced by defects, we first note that the atomic scale defects discussed here are not charged defects.  
This can be shown by the scanning tunneling spectroscopy measurements~\cite{SKSong2019a}, which showed that the above atomic scale defects do not shift the energy of the spectral features in their vicinity.   
Therefore, in order to understand the role of the defects in EPS, it is necessary to establish the detailed structural relationship between the electronic domain boundaries and the defects.
As already mentioned, most of the PFDs are located at the domain boundaries between $4\times1$ and $8\times2$ phases or within the metallic $4\times1$ domains.
In Fig.~\ref{fig2}(a), we close-up a single metallic $4\times1$ domain to reveal the detailed local structures of domain boundaries.
One can confirm that most of the defects involved are the strong PFDs. 
In particular, we mark the CDW orientations as pinned by the defects with red and blue bars in Fig.~\ref{fig2}(a) and the red and blue stripes in Fig.~\ref{fig2}(b). 
The length scale of the constraint by a PFD along the wire is assumed from the decay lengths of the local lattice distortion induced by the absorbates~\cite{DOh2014a,SYu2006a} and the coherence length of the CDW~\cite{Kim2012n}, $\sim 10a_0$ ($a_0$ is the lattice constant of the substrate surface, $0.384~nm$). 
The color plot of Fig.~\ref{fig2}(b) clearly differentiates the characteristics of the insulating and metallic areas. 
The insulating phase persists away from the PFDs (top left of Figs.~\ref{fig2}) and where the constraints by the PFDs obey properly the alternating CDW orientations between neighboring wires (top right and bottom middle of Figs.~\ref{fig2}).
There exist few exceptional wires in this rule.
For these wires, the chiral solitons are trapped to fix the wrong orientation of CDW imposed by particular defects~\cite{Kim2012n} [the yellow box in Fig. 2(a)]. 
The alternating CDW orientations between neighboring wires is the prerequisite for the insulating $8\times2$ 2D CDW ordering. 
In sharp contrast, the intra- and inter-wire CDW orders were obviously broken in the metallic domain. 
The broken intrawire order is shown by the overlapped color and the interwire disorder is represented by the same colors repeating in neighboring wires. 
These areas with the frustrated CDW order show either the $4\times1$ structure [Fig.~\ref{fig3}(c)] or a characteristic $4\times2$ structure [Fig.~\ref{fig3}(d)].
The latter is a different $\times2$ structural distortion from the insulating wires, which was observed in the previous STM studies around defects~\cite{SJPark2004a}. This structure is metallic and its STM images are consistent with the previous DFT calculations for the trimer distortion with metallic bands~\cite{JHCho2001a,JHCho2005a}.
The previous DFT calculation~\cite{Stekolnikov2007} further indicated that the $4\times2$ trimer structure is energetically competing with the metallic $4\times1$ structure but is less stable than the $8\times2$ CDW structure [Fig.~\ref{fig3}(e)].
In turn, it suggests that the constraints on the CDW local order by the defects prevent the $\times2$ order along the wire or its interwire $\times8$ order to destabilize the $8\times2$ structure in favor of the $4\times1$ or the $4\times2$ structures, respectively.


The above phenomenological rule of the domain boundary can further be confirmed at different temperatures. 
The STM images of domain boundaries formed at a higher temperature of 95 K are shown in Fig.~\ref{fig3} and Fig.~S5.
In contrast to 78~K, most part of the surface is held in the metallic phase, while only small insulating $8\times2$ domains are observed.
Note that the strong PFDs are tightly bound to the domain boundaries for those small insulating domains.
We can confirm that the PFDs are arranged to keep the alternating CDW orientation of the surviving $8\times2$ domains in Fig.~\ref{fig3}(a). 
This may be compared with what was observed for oxygen impurities for the early CDW condensation above the transition temperature, where the distance between neighboring defects within a wire (being commensurate to the CDW periodicity or not) was the crucial parameter~\cite{HYeom2016b}. 
The commensurate distance between neighboring defects is also naturally requested in the present case.
The distorted metallic $4\times2$ structures are also observed nearby the defects. 
In another small insulating domain shown in Fig.~\ref{fig3}(b), one can reconfirm the same rule of the coherent overlap of the CDW constraint between the pinning defects for the insulating domain.  
In addition, we observe a defect free wire with a CDW segment between the wires with the strong PFDs [arrowed in Fig.~\ref{fig3}(b)]. 
We believe that this CDW segment is caused by the substantial interwire coupling of the CDW, which is natural for the $8\times2$ structure.  

\begin{figure}[tb]
\includegraphics[width=8cm]{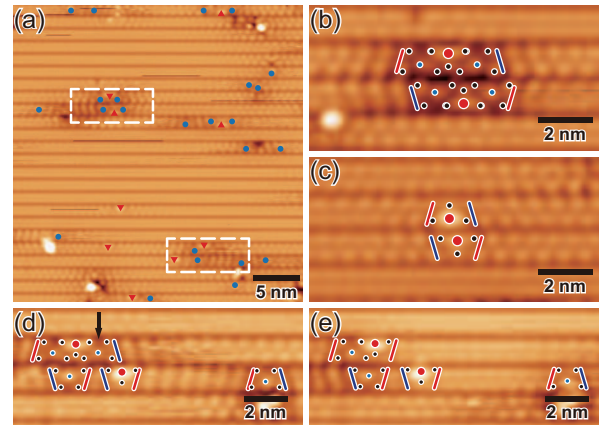}%
\caption{
(a) STM topographic images ($V_s=-0.5$~V) of the electronic phase separation at 120~K slightly below the pristine CDW transition temperature of 130 K.
White boxes indicate some of local $8\times2$ patches formed near defect clusters.
Blue dots and red triangles indicate typical PFDs in Figs.~\ref{fig1}(b) and (d).
(b), (c) Close-up STM images showing very short $8\times2$ or $4\times2$ structures decaying rapidly from defect clusters.
(d), (e) Two consecutive STM images of a short $8\times2$ structure formed in between two neighboring PFDs.
}
\label{fig4}
\end{figure}

The domain boundary formation mechanism revealed above manifests itself in a dynamical but consistent way at a even higher temperature of 120~K, close to the pristine transition temperature.
At 120 K, it is hard to find an insulating $8\times2$ domain, but only very short segments mostly  with the $4\times2$ structure are found to be pinned to clusters of defects [Figs.~\ref{fig4}(a), (b), and (c)].
Short CDW segments occur when two neighboring defects within a single wire have proper CDW constraints and there is at least another defect in the neighboring wire to guarantee locally the $8\times2$ structure [Fig.~\ref{fig4}(d)].
The strong PFDs become mobile at this temperature as already reported~\cite{SJPark2005a} (see alss supplementary materials Fig.~S4).
For the particular defect clusters in Fig.~\ref{fig4}(d), one strong PFD [arrowed in Fig.~\ref{fig4}(d)] is detrapped to disappear from the image in the next moment and the local CDW distortion vanishes for that wire and also for the neighboring wire [Fig.~\ref{fig4}(e)]. 
This event and a few similar ones observed corroborate the two important factors to maintain a local CDW patches or domains, namely, the coherent overlap of the CDW phase constraints from neighboring defects and the interwire coupling, as discussed above.


In summary, we have investigated the atomic scale domain boundaries formed on the electronically phase separated In/Si(111) surface at various temperatures.
Atomically-resolved STM images reveal characteristic atomic scale defects, which induce the CDW phase flip next to the defects.
The major type of defects originates from the extra In adatoms and the domain boundaries are largely decided by this type of defects. 
The detailed analysis of STM images reveals further that the constraints of the CDW phase imposed by the defects break the coherent CDW ordering to induce local metallic domains.  
At a temperature close to the transition temperature, the motion of the defects is activated and induces the dynamical EPS in a very short length scale but its mechanism is consistent with that at a lower temperature.
This result phenomenologically unveils the atomistic origin of the EPS in a CDW system as induced by atomic scale defects collectively, emphasizing the importance of the atomic scale approach to understand inhomogeneous electronic phases.
Defect engineering would be a promising way to manipulate phase-separated electronic systems. 

This work is supported by Institute for Basic Science (Grant No. IBS-R014-D1).



%

\end{document}